\documentclass[pra,letterpaper,twocolumn,showpacs,superscriptaddress,floatfix]{revtex4}
\usepackage{graphicx,psfrag,amsmath,amssymb,amsfonts,bbm,latexsym,color,dcolumn,bm}

\begin{document}

\title{Thermodynamical approaches to efficient sympathetic cooling \\ 
in ultracold Fermi-Bose atomic mixtures}

\author{Michael Brown-Hayes} 

\affiliation{Department of Physics and Astronomy,Dartmouth College,6127 Wilder Laboratory,Hanover,NH 03755}

\author{Qun Wei} 

\affiliation{Department of Physics and Astronomy,Dartmouth College,6127 Wilder Laboratory,Hanover,NH 03755}

\author{Carlo Presilla} 

\affiliation{Dipartimento di Fisica,Universit\`a di Roma "La Sapienza",Piazzale A. Moro 2,Roma 00185,Italy} 
\affiliation{Center for Statistical Mechanics and Complexity,INFM-CNR,Unit\`a di Roma 1,Roma 00185,Italy} 
\affiliation{INFN, Sezione di Roma 1, Roma 00185, Italy}

\author{Roberto Onofrio} 

\affiliation{Dipartimento di Fisica ``Galileo Galilei'',Universit\`a di Padova,Via Marzolo 8,Padova 35131,Italy}
\affiliation{Center for Statistical Mechanics and Complexity,INFM-CNR,Unit\`a di Roma 1,Roma 00185,Italy}
\affiliation{Department of Physics and Astronomy,Dartmouth College,6127 Wilder Laboratory,Hanover,NH 03755} 

\date{\today}

\begin{abstract}
We discuss the cooling efficiency of ultracold Fermi-Bose mixtures in 
species-selective traps using a thermodynamical approach.  
The dynamics of evaporative cooling trajectories is analyzed in the 
specific case of bichromatic optical dipole traps also taking into 
account the effect of partial spatial overlap between the Fermi 
gas and the thermal component of the Bose gas. 
We show that large trapping frequency ratios between the Fermi and the
Bose species allow for the achievement of a deeper Fermi degeneracy, 
consolidating in a thermodynamic setting earlier arguments based on 
more restrictive assumptions. In particular, we confirm that the 
minimum temperature of the mixture is obtained at the crossover 
between boson and fermion heat capacities, and that below such 
a temperature sympathetic cooling vanishes. 
When the effect of partial overlap is taken into account, optimal 
sympathetic cooling of the Fermi species may be achieved by properly 
tuning the relative trapping strength of the two species in a time-dependent fashion. 
Alternatively, the dimensionality of the trap in the final stage of
cooling can be changed by increasing the confinement strength, which 
also results in a crossover of the heat capacities at deeper Fermi 
degeneracies. 
This technique may be extended to Fermi-Bose degenerate mixtures in optical lattices.
\end{abstract}


\maketitle

\section{Introduction}  
Atomic physics and condensed matter physics now enjoy strong
connections through the study of quantum transport in ultracold dilute
gases \cite{Reviewsbooks}. Long-standing problems of condensed matter 
physics may be addressed by preparing controllable environments for 
the dynamics of cold atoms and by continuously tuning their interactions.  
This in turn allows for the study of fundamental features of high-temperature 
superconductivity using ultracold gases as controllable, analog 
computers of various model hamiltonians \cite{Levin}.  

Degenerate Fermi gases were first produced in 1999 \cite{Jin}, and 
more recently Fermi superfluid behaviour has been conclusively
evidenced through the generation of vortices \cite{Zwierlein1} and 
the onset of critical velocities \cite{Miller} in degenerate samples
of $^6$Li. Weakly interacting Fermi gases are difficult to bring 
to quantum degeneracy mainly due to fundamental obstacles in adapting 
cooling techniques successfully used for bosonic species.  
In particular, the Pauli principle inhibits efficient evaporative
cooling among identical fermions as they reach degeneracy.  
This issue has been circumvented by developing two cooling techniques,
namely mutual evaporative cooling of fermions prepared in two
different states and sympathetic cooling with a Bose species.  
In the case of dual evaporative cooling, a selective removal of the
most energetic fermions in both the hyperfine states is performed.  
Provided that the initial number of atoms in each state is roughly the
same, efficient dual evaporative cooling can be performed throughout the entire process. 
Limits to the minimum reachable absolute temperature using dual evaporative
cooling have been addressed in \cite{Crescimanno}, resulting in a
minimum reachable temperature $T \simeq \mu/k_B$, with $\mu$ the 
chemical potential of the Fermi gas (see also \cite{Holland} for 
a complementary analysis). Moreover, the number of 
available atoms $N_f$ progressively decreases over time with 
a corresponding drop in the Fermi temperature $T_{\mathrm F}$ proportional 
to $N_{f}^{1/3}$. The resulting gain in terms of a lower $T/T_F$ 
degeneracy ratio is therefore limited, and the smaller clouds obtained 
at the end of the evaporative cooling are detrimental to 
detailed experimental investigations requiring a large number 
of atoms, such as a quantitative mapping of the superfluid phases. 
In the case of sympathetic cooling using a Bose gas, the number 
of fermions is instead kept constant, leaving aside losses due to 
background pressure and two- and three-body collisions, and the 
cooling efficiency depends on the optimization of Fermi and Bose 
collisional properties, heat capacities, and, in the case 
of inhomogeneous samples, their spatial overlap. 

To date, the smallest Fermi degeneracy achieved with both cooling 
techniques is in the $T/T_{\mathrm F } \geq 5 \times 10^{-2}$ range 
\cite{Hadzibabic,Grimm}. This limitation has not precluded the study 
of temperature-independent features of degenerate Fermi gases, such 
as quantum phase transitions related to unbalanced spin populations 
\cite{Zwierlein2,Partridge,Zwierlein3,Shin} or the effect of Fermi 
impurities in the coherence properties of a Bose gas \cite{Gunther,Ospelkaus}.  
However, the study of more conventional phase transitions in which the 
temperature is the key parameter is still uncharted territory and, as 
discussed for instance in \cite{Chen,Bulgac,Chien}, this requires the 
achievement of degeneracy factors $T/T_{\mathrm F} \simeq 10^{-3}$ or lower.  
Unconventional pairing mechanisms that are unstable at higher 
$T/T_{\mathrm F}$ could then be observed, and the phase diagram 
of Fermi atoms in the degenerate regime could be mapped in a wider 
range of parameter space.
Moreover, the study of ultracold Fermi-Bose mixtures is an interesting
subject in itself, acting as the counterpart of the $^3$He-$^4$He
liquid mixtures extensively investigated at much higher densities 
and temperatures. 

Considering the novel physical insights that deeper Fermi degenerate
gases and Fermi-Bose mixtures may provide, it is relevant to discuss 
the limitations to reaching the lowest $T/T_F$ in realistic settings 
available by means of sympathetic cooling, and ways to overcome them.
Here, we discuss two different techniques to overcome the apparent 
$T/T_F \simeq 10^{-2}$ limit observed so far, based on optimized 
heat capacity matching with species-selective traps or with 
lower dimensionality traps. The paper is organized as follows. 
In Section II, after briefly reviewing previous results on sympathetic 
cooling in species-selective traps, we determine the time evolution 
of the temperature of the mixture in a particular class of
species-selective traps through a thermodynamical analysis, and 
we subsequently 
include the effect of the spatial overlap between the thermal
component of the Bose gas and the Fermi gas. 
The main novelty with respect to previous semiquantitative analyses 
is that including both temporal and spatial dependence in the
thermodynamics of sympathetic cooling leads to an optimization 
of the heat capacity provided that a time-dependent trapping 
frequency ratio is implemented. 
As an alternative to this optimization procedure, in Section III 
we discuss heat capacity matching resulting from lower effective 
dimensions for trapping, exploiting the strong dependence on 
dimensionality of the density of states of the Bose gas. 
A simpler protocol for optimizing heat capacity is available with 
nearly one-dimensional Fermi-Bose mixtures, which requires time 
modulation of the trapping strengths of the Fermi and 
Bose gases in the last cooling stage.  
Broader considerations on generic trapping settings are then discussed 
in the conclusions.

\section{Heat capacity matching through species-selective trapping}

After discussing the evidence for a correlation between the degeneracy
factor $T/T_F$ and the trapping frequency ratio between the Fermi and
the Bose species, we review previous results on the use of
species-selective traps and their limiting assumptions. We then 
relax these assumptions with a thermodynamical analysis also 
including the effect of the partial overlap between the Fermi gas 
and the thermal component of the Bose gas in the specific case of 
bichromatic traps.

\subsection{Qualitative considerations on heat capacity matching} 

Evaporative cooling has been instrumental in reaching Bose degeneracy 
for dilute atomic gases.  Extensive analysis have already addressed 
the dynamics of evaporative cooling of a Bose gas 
\cite{Hess,Luiten,Holland1,Pakarinen}, using Monte-Carlo 
\cite{Wu,Arimondo}, mean-field analysis \cite{Avdeenkov},
and beyond \cite{Olsen}, including also more detailed effects
\cite{Yamashita,Tol,Comparat}.  
These studies have been also extended to the case of separate Bose and
Fermi clouds \cite{Butts} or Fermi-Bose mixtures using the quantum
Boltzmann equation \cite{Geist} or other semiclassical models \cite{Carr}.
Some generic features of the effectiveness in cooling fermions
through a Bose gas can be addressed based on the insights first 
discussed in \cite{Truscott} and then analyzed in more detail in
\cite{Presilla,OnofrioJSP}. 
The heat capacity of a degenerate Fermi gas depends linearly on its 
temperature, being for a harmonically trapped gas equal to 
$C_{f} \simeq \pi^2 k_B N_{f} T/T_F$, while a harmonically 
trapped degenerate Bose gas has a cubic dependence on 
temperature $C_{b} \simeq 10.8 ~k_B N_{b} (T/T_c)^3$.  
The degeneracy parameter can be written in terms of 
the ratio of heat capacities: 
$T/T_F \simeq 0.35 (\omega_{b}/\omega_{f})^{3/2}
(C_{b}/C_{f})^{1/2}$ \cite{OnofrioJSP}.  
By assuming that sympathetic cooling loses efficiency when the 
heat capacity of the Bose gas matches exactly that of the Fermi
gas ($C_{b} =C_{f}$) we obtain a conservative limit on the 
attainable $T/T_F$ {\it vs.} $\omega_{f}/\omega_{b}$ space, depicted 
in Fig. 1 by the upper line.  In the hypothesis that some residual 
cooling occurs when $C_{b}<C_{f}$, for 
instance with cooling stopping when $C_{b}/C_{f} \leq 0.1$, we obtain 
the lower line in Fig. 1. Realistically, we do expect that sympathetic cooling 
will be quenched when $0.1 \leq C_{b}/C_{f} \simeq 1$, {\it i.e.} 
in the region delimited by the two lines. 
In Fig. 1 we also plot the minimum $T/T_F$ as obtained 
by the various running experiments with Fermi-Bose mixtures. 
Although the number of explored Fermi-Bose mixtures is limited 
and the diverse technical solutions for trapping and cooling 
may provide alternative explanations \cite{note1}, a correlation 
between the trapping frequency ratio $\omega_{f}/\omega_{b}$ and the 
minimum achieved $T/T_F$ seems corroborated by the actual 
results and invites more quantitative attention.
\begin{figure}
\includegraphics[width=1.00\columnwidth,clip]{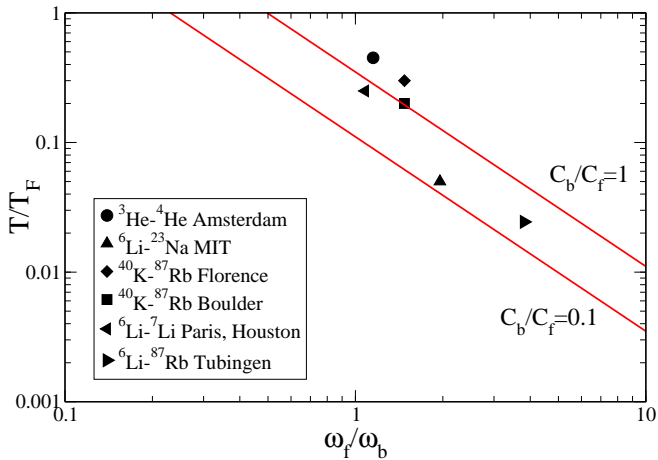}
\caption{(Color online) Plot of experimentally obtained $T/T_F$ {\it versus} 
the trapping frequency ratio $\omega_f/\omega_b$.  
Lines indicate the theoretically predicted range of $T/T_F$ 
values based on a semiquantitative heat capacity matching argument 
as discussed in the text. The experimental data are taken from 
\cite{McNamara} (Amsterdam), \cite{Hadzibabic} (MIT), 
\cite{Ferlaino} (Florence), \cite{Goldwin} (Boulder), 
\cite{Truscott} (Houston), \cite{Schreck} (Paris),
 \cite{Silber} (Tubingen).}
\label{ttfexperiments}
\end{figure}
Further analysis provided insight into the cooling limitations in
different trapping conditions and for different species combinations
\cite{BrownHayes}.  It was found that significant gains in
$T/T_F$ could be achieved for stronger relative Fermi-Bose
confinements than the {\sl natural} $\omega_{f}/\omega_{b}$ 
provided by the mass ratio between the two species. 
In \cite{BrownHayes} the focus was on an equilibrium situation at 
nearly zero temperature, and to develop a more comprehensive understanding 
of the cooling process a dynamical framework is needed in which crucial finite 
temperature effects for the Bose gas are taken into account. 
Here we specialize the analysis of the thermodynamics as 
${}^{6}$Li is sympathetically cooled by evaporating ${}^{87}$Rb 
in an optical dipole trap. This mixture, expected to optimize 
cooling efficiency \cite{BrownHayes,Eddy}, is currently used in various 
laboratories \cite{Silber,LaserP,Taglieber} both for studies of
Fermi superfluidity and for the formation of ultracold molecules
\cite{Gonzalez}, taking advantage of the large electric dipole moment of 
Li-Rb molecules \cite{Aymar}. 
 
In all previous discussions of species-selective traps a static picture was 
assumed for the cooling of the Fermi-Bose mixture, relying on the analysis 
already available in the case of single-species trapping \cite{Ohara}.
A critical element of this cooling is to balance the thermalization
and loss rates; rapid trap modification allows for minimization of atom losses 
but is limited by the requirement that we go proceed slowly enough to keep the 
system at thermal equilibrium.  It has been shown
\cite{VanDruten} that thermodynamic equilibrium is maintained if the
ratio between the potential depth and the atomic cloud temperature,
given by $\eta \equiv \Delta U/k_{B}T$, is kept constant.
The time-dependence of the potential depth is then given by \cite{Ohara}:
\begin{equation}
\frac{\Delta U(t)}{\Delta U_{i}} = 
\left(1+\frac{t}{\tau}\right)^{\eta_{e}}
\label{eqn1}
\end{equation}
where $\Delta U_{i}$ is the initial potential depth, $\eta_{e}=-2(\eta'-3)/\eta'$, and
$\tau^{-1}=(2/3)\eta'(\eta-4)\exp(-\eta)\gamma_i$, with
$\eta'=\eta+(\eta-5)/(\eta-4)$ and $\gamma_i$ the initial elastic
collision rate.  With the time dependence of $\Delta U$ determined 
in this way we can obtain the other relevant quantities in the process
(number of particles, temperature, density, scattering rates)
resulting in scaling laws similar to Eq. (\ref{eqn1}). 
A potential depth/temperature ratio of $\eta$=5 to 10 is considered 
to yield optimal efficiency in evaporative cooling \cite{VanDruten}.
This approach can be applied to either the simple case of evaporative 
cooling of a single Bose species, or the sympathetic cooling in a 
Fermi-Bose mixture.  In both cases, Eq. (\ref{eqn1}) describes the 
potential depth of the Bose species $\Delta U_{b}$, with the implicit 
assumption that the presence of the fermions does not drastically 
affect the evaporation and cooling of the bosons.   This is justified since all 
dual-species systems trap at least an order of magnitude more bosons than 
fermions, but this approximation may suffer towards the end of the cooling 
process when a majority of the bosons have been evaporated.

In the case of a bichromatic optical dipole trap, from $\Delta U_b$ we
determine the required power $P_1$ of the laser confining both species, and
then for a targeted trapping strength ratio $\omega_{f}/\omega_{b}$ we
can calculate the required power $P_2$ for the Bose-deconfining laser, 
and finally determine the fermion potential depth $\Delta U_{f}$.  
In this way we have independent control of the spatial size and 
potential depth for each of the two species, allowing us to either maintain 
a constant $\omega_{f}/\omega_{b}$ throughout cooling or adjust 
the relative trapping strengths during the process.  

The exact way in which the temperature is determined from the trapping
parameters depends upon both the model and the proposed cooling
strategy. Eq. (\ref{eqn1}) merely identifies a limit to the cooling
rate if thermodynamic equilibrium is to be maintained throughout the 
process, and assuming a constant ratio $\eta$, as it was discussed 
in \cite{Presilla}. In practice, these assumptions can be relaxed 
by using a more dynamical model based on energy balance, as we 
discuss in the next section.  

\subsection{Thermodynamical balance} 

To move beyond the limiting assumptions in \cite{Presilla}, we start our 
analysis by considering a dual-species system at thermodynamic 
equilibrium. The trap parameters are then suddenly changed in order 
to force some bosons to evaporate; then one waits for a new 
thermodynamic equilibrium before applying another evaporation step, 
in analogy to the scheme discussed in \cite{Davis}. 
The step-by-step temperature reached in this way is determined 
by energy conservation for the Fermi and Bose gases \cite{Wouters}.

For concreteness, we consider a species-selective trapping scheme as 
described in \cite{Onofrio}. A mixture of $N_f$ fermions and $N_b$
bosons is confined into a bichromatic optical dipole trap tailored 
by two lasers of wavelengths $\lambda_1$, $\lambda_2$ and powers $P_1$, $P_2$.
In order to obtain quasi analytic results, we will approximate 
the trap potential by a truncated harmonic potential properly 
reproducing the bottom curvature and the depth of the well.
This approximation becomes exact for energies small with respect 
to the trap depth, a condition satisfied in the cases discussed below. 
The presence of the second laser allows one to make the trap parameters
of the fermion species, namely the characteristic frequencies 
$\omega_{fx},\omega_{fy},\omega_{fz}$ and the depth $\Delta U_f$, 
different from the corresponding boson parameters,
$\omega_{bx},\omega_{by},\omega_{bz}$ and $\Delta U_b$.
As a consequence, the ratio $\omega_f/\omega_b$, for each species $s=b,f$ 
we define $\omega_{s}=(\omega_{sx} \omega_{sy} \omega_{sz})^{1/3}$,
can be varied from its mass-determined value for $P_2/P_1=0$ to an ideally
arbitrary large value for $P_2/P_1$ approaching a positive critical value 
\cite{Onofrio}.
The evaporation steps are carried out by decreasing the power $P_1$ of the
reference laser while maintaining the ratio $P_2/P_1$ at a constant value.
In this way, the trap depths $\Delta U_b$, $\Delta U_f$, 
which are proportional to $P_1$, and the frequencies $\omega_b$, $\omega_f$ 
(proportional to $\sqrt{P_1}$), decrease 
while the ratio $\omega_f/\omega_b$ remains constant \cite{note2}. 

At the end of an evaporation step in which the power of the reference laser
is changed from $P_1$ to $P_1+dP_1$ and once thermodynamic equilibrium 
is reestablished, the temperature changes from $T$ to $T+dT$ according to
an energy balance equation of the form
\begin{equation}
\left(\Delta U_b +\xi k_BT\right) dN_b^{ex} = dE_b + dE_{f} {\ \ ,}
\label{energybalance}
\end{equation}
where $dN_b^{ex}$ is the number of bosons in the excited states 
at temperature $T$ that evaporate and $\Delta U_b +\xi k_BT$ 
(with $0 \leq \xi \leq 1$) is the mean 
energy per evaporated boson \cite{Luiten}.   The 
quantities $dE_b$ and $dE_f$ are the energy changes of the 
trapped boson and fermion species, both at temperature $T$.

By observing that due to a change of $P_1$, all the quantities 
$T$, $\omega_b$, $\Delta U_b$, $\omega_f$, $\Delta U_f$ change
and using for $N_b^{ex}$, $E_b$ and $E_f$ the expressions provided 
by Eqs. (\ref{Nb4}), (\ref{Eb4}) and (\ref{Ef2}), we have

\begin{eqnarray}
dN_b^{ex} &=& 
\frac{\partial N_b^{ex}}{\partial T}dT
+\frac{\partial N_b^{ex}}{\partial \omega_b} d\omega_b
+\frac{\partial N_b^{ex}}{\partial \Delta U_b}d \Delta U_b {\ ,} 
\label{dNb}
\\
dE_b &=& 
\frac{\partial E_b}{\partial T}dT
+\frac{\partial E_b}{\partial \omega_b} d\omega_b
+\frac{\partial E_b}{\partial \Delta U_b}d \Delta U_b {\ ,} 
\label{dEb}
\\
dE_f &=&
\frac{\partial E_f}{\partial T}dT
+\frac{\partial E_f}{\partial \omega_f}d \omega_f { \ .}
\label{dEf}
\end{eqnarray}

\noindent
It is worth to point out that, within the approximation used, $E_f$ does not depend
on $\Delta U_f$ and the number of fermions $N_f$ remains constant.
By evaluating the above partial derivatives and inserting the result
into Eq. (\ref{energybalance}), we arrive at 
\begin{widetext}
\begin{eqnarray}
\label{ttfp1ode}
\frac{d \frac{T}{T_F}}{d P_1} = 
\frac{\frac{T}{T_F}}{2 P_1} ~
\frac{\frac{3}{\pi^2}\left(\frac{\omega_f}{\omega_b}\right)^3 
\left(\frac{T}{T_F}\right)^2
p\left(\frac{\Delta U_b P_1^{-1/2}}{k_B T_F }
\left(\frac{T}{T_F}\right)^{-1} P_1^{1/2} \right) + \frac{1}{2} 
+ \frac{3}{4\pi^2}\left(\frac{T}{T_F}\right)^{-2}}
{\frac{3}{\pi^2}\left(\frac{\omega_f}{\omega_b}\right)^3 
\left(\frac{T}{T_F}\right)^2 q\left(\frac{\Delta U_b P_1^{-1/2}}{k_B T_F } 
\left(\frac{T}{T_F}\right)^{-1} P_1^{1/2} \right) -1 } {\ ,}
\end{eqnarray}
\end{widetext}
where we used $d (T/T_F) = d T/T_F - (T/T_F) dT_F/T_F$ and $dT_F/T_F =
dP_1/(2P_1)$, which stems from the proportionality of $T_F$ 
to $\omega_f$, namely  $k_B T_F=(6 N_f)^{1/3} \hbar \omega_f$, see Eq. (\ref{Nf2}). 
Like $\omega_f/\omega_b$, the value of $\Delta U_b P_1^{-1/2} / k_B T_F$
is a constant determined by the value of the ratio $P_2/P_1$, and $p(x)$ and $q(x)$ are defined as
\begin{eqnarray}
p(x) &=& \int_0^x \frac{t^3}{e^t-1} dt + \xi \frac{x^3}{e^x-1}, \\
q(x) &=& 3 x \int_0^x \frac{t^2}{e^t-1} dt - 4 \int_0^x \frac{t^3}{e^t-1} dt
\nonumber
\\
&&+3 \xi \int_0^x \frac{t^2}{e^t-1} dt - \xi \frac{x^3}{e^x-1} {\ .}
\label{pdef}
\end{eqnarray}
Note that $p(0)=q(0)=0$ whereas $p(x) \simeq 6 \zeta(4)$ and 
$q(x) \simeq  6 \zeta(3) x -24 \zeta(4)+6\zeta(3)\xi$ for $x\gg 1$, 
$\zeta$ being the Riemann zeta function.

\begin{figure}[t]
\includegraphics[width=1.00\columnwidth,clip]{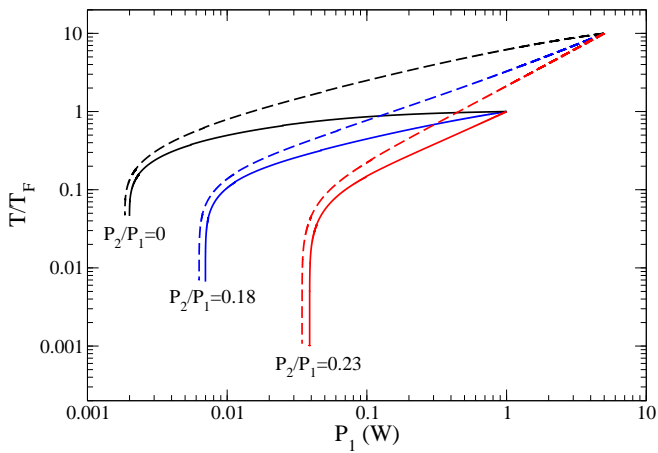}
\caption{(Color online) Dependence of the degeneracy factor $T/T_F$ upon 
the confining laser power $P_1$ during sympathetic forced 
evaporative cooling as determined by Eq. (\ref{ttfp1ode}).  
The system is a mixture with $N_f$ atoms of $^6$Li and $N_b$ atoms 
of $^{87}$Rb trapped in a bichromatic optical dipole trap shaped by two lasers 
of power $P_1$ and $P_2$ at the wavelengths of $\lambda_1=$ 1064 nm
and $\lambda_2=$ 740 nm for the $^6$Li-$^{87}$Rb mixture as chosen 
in \cite{OnofrioJSP}.  Two sets of curves are shown for different 
initial conditions and for different values of the ratio $P_2/P_1$,
kept constant during the evaporation.  The fermion-to-boson trapping 
frequency ratio, determined by $P_2/P_1$, is $\omega_f/\omega_b=2.443$ for $P_2/P_1=0$,
$\omega_f/\omega_b=8.186$ for $P_2/P_1=0.18$ and
$\omega_f/\omega_b=15.911$ for $P_2/P_1=0.23$.  
For the same $P_2/P_1$ values, the other constant $\Delta U_b P_1^{-1/2} / k_B T_F$ 
which appears in Eq. (\ref{ttfp1ode}) amounts to 18.73, 3.70 and 0.59,
respectively.  For simplicity, we set $\xi=0$. We assume $N_f=10^4$ is constant during the evaporation
and, for the continuous (dashed) curves the initial number of
bosons is $N_b=2\times 10^7$ ($N_b=2\times 10^9$), Fermi
degeneracy $T/T_F=1$ ($T/T_F=10$) and $P_1=1~W$ ($P_1=5~W$).  
The minimum achievable $T/T_F$, corresponding mathematically
to a singularity of Eq. (\ref{ttfp1ode}) and physically to a fermion-boson heat
capacity equality, (a) does not depend on $N_f$ and $N_b$, provided that
$N_b$ is sufficiently large and (b) depends only slightly on the initial
conditions for $T/T_F$ and $P_1$, but (c) decreases appreciably if the 
trapping frequency ratio $\omega_f/\omega_b$ is increased.}
\end{figure}
\begin{figure}[t]
\includegraphics[width=1.00\columnwidth,clip]{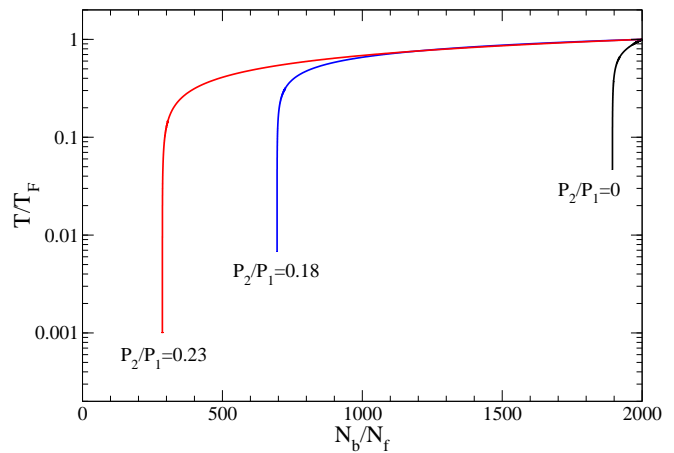}
\caption{(Color online) Dependence of the degeneracy factor $T/T_F$ upon the number of bosons
(normalized to the number of fermions) during a sympathetic 
forced evaporative cooling driven by the laser power $P_1$ as 
determined by Eq. (\ref{ttfp1ode}). The parameters are 
the same as in the case of the continuous curves in 
Fig. 2, with the initial values of $N_b=2 \times 10^7$, $T/T_F=1$, and $P_1=1~W$.
It is evident that the use of larger $\omega_f/\omega_b$ ratios 
allows us to reach a deeper Fermi degenerate regime, which amounts 
to a gain by almost two orders of magnitude difference in the case 
of the larger $\omega_f/\omega_b$ ratio.}
\end{figure}

Equation (\ref{ttfp1ode}) is a nonlinear ordinary differential equation 
which allows for the determination of $T/T_F(P_1)$ during the evaporative cooling.
Observing that both functions $p(x)$ and $q(x)$ are non negative for $x \ge 0$
and assuming for simplicity $\xi=0$, the qualitative behavior of 
$T/T_F(P_1)$ is as follows.
The numerator of the last fraction in Eq. (\ref{ttfp1ode}) is always positive.
If we start from initial values of $P_1$ and $T/T_F$ respectively 
not too small and not too large, the argument of the functions $p$ and
$q$ is large with respect to unity, which is equivalent to state that 
$\Delta U_b \gg k_B T$, a fact that also justifies the choice $\xi=0$.
The denominator of the last fraction in Eq. (\ref{ttfp1ode}) is thus also
positive so that $T/T_F$ decreases by decreasing $P_1$. 
The decrease may be faster or slower than $P_1^{1/2}$ depending on 
the value of the constants $\omega_f/\omega_b$ and $\Delta U_b P_1^{-1/2} / k_B T_F$.
Eventually, however, the last fraction in Eq. (\ref{ttfp1ode}) becomes 
larger than 1 so that a second regime starts in which $T/T_F$ decreases faster and faster.
As a consequence, the argument of the functions $p$ and $q$ decreases
and the denominator of the last fraction in Eq. (\ref{ttfp1ode}) approaches 0.
A singular point is thus reached in which $d(T/T_F)/dP_1=\infty$ and $T/T_F>0$. 
A numerical study also shows that Eq. (\ref{ttfp1ode}) has a
discontinuity at the singular point with unphysical negative values of 
$T/T_F$ on the left. 
The value of $T/T_F$ at the right of the singular point represents the minimum achievable $T/T_F$ 
during the cooling, provided that the initial number of bosons is
sufficiently large so that they are not completely evaporated before the singular point is reached.

The behavior of $T/T_F$ as a function of $P_1$ is shown in
Fig. 2 for different initial conditions and different values
of $\omega_f/\omega_b$.
We stress that whereas the minimum $T/T_F$ depends very little
on the details of the cooling, the number of atoms of both species and initial
values of temperature and reference laser power, we observe a
substantial decrease of the minimum achievable $T/T_F$, by increasing the 
ratio $\omega_f/\omega_b$.
This is in agreement with a previous prediction based on a rough matching 
of boson and fermion heat capacities \cite{Presilla}.
In fact, the singular point of Eq. (\ref{ttfp1ode}) is defined by the condition
\begin{eqnarray}
\left(\Delta U_b +\xi k_BT\right)
\frac{\partial N_b^{ex}}{\partial T} -
\frac{\partial E_b}{\partial T} -
\frac{\partial E_f}{\partial T} = 0 {\ .}
\label{singularpoint1}
\end{eqnarray}
\noindent
For $\Delta U_b \gg k_B T$, a condition well satisfied at the singular point, we have 
\begin{eqnarray}
\frac{\partial N_b^{ex}}{\partial T} =
\frac{1}{k_B T}
\frac{\partial E_b}{\partial T} 
\frac{6\zeta(3)}{24\zeta(4)}_{\ ,}
\end{eqnarray}
therefore Eq. (\ref{singularpoint1}) is equivalent to
\begin{eqnarray}
\left(0.28 \frac{\Delta U_b}{k_BT} -1 \right) C_b \simeq C_f {\ ,}
\label{singularpoint2}
\end{eqnarray}
where $C_s=\partial E_s/\partial T$ with $s=b,f$.

The sharp drop observed in $T/T_F$ before the singular point
deserves some comments. At temperatures sufficiently low with respect to $T_F$ and $T_c$, 
the energy of the mixture $E_b+E_f$ is dominated by the zero temperature
Fermi energy $E_f(N_f,0)$, see Appendix A. In this case, the
right-hand side of the energy balance equation (\ref{energybalance}) 
can be approximated by $(\partial E_f(N_f,0)/\partial \omega_f) d \omega_f$, and 
in the proximity of the singular point (i.e. at low temperatures), 
following Eq. (\ref{singularpoint2}) the left-hand side of the same balance can be written as 
$(\gamma C_b - C_f) dT$, where $\gamma \simeq 0.28 \Delta U_b/k_BT -1$.
The approximated energy balance thus gives
\begin{equation}
  \label{large-dT}
  d T \simeq \frac{d \omega_f}{\omega_f} 
\frac{\frac{3}{4} N_f k_B T_F}{\gamma C_b - C_f}_{\ .}
\end{equation}
From Eq. (\ref{large-dT}) we see that a small decrease of the 
fermion trapping frequency induces a temperature decrease, the size of which 
depends on the value of $N_f k_B T_F/(\gamma C_b - C_f)$.
Note that the denominator of this ratio contains a difference, not a sum, 
of the boson and fermion specific heats.
The divergence of the derivative $dT/d\omega_f$ predicted at the critical 
point is certainly unphysical:  we expect that a long time is needed 
to re-equilibrate the system in a freezing step  $T \to T+dT$ with $dT$ large.
In this case, dissipative phenomena should be taken into account 
by a more complicated model in which the singular point will be substituted 
by a minimum. However, this does not change the meaning of the lowest reachable $T/T_F$
which is the point where fermion and boson specific heats do match.

In Fig. 3 we plot the dependence of $T/T_F$ upon the number of bosons,
normalized to the fermion number (assumed to be constant during the
evaporation process). It is manifest that deeper
Fermi degeneracy factors are obtained for higher trapping frequency 
ratios. This plot has to be compared to the one presented in 
Fig. 2 of \cite{Carr} in which $T/T_F$ was shown versus 
a similar quantity (in our notation $(N_b^{(0)}-N_b)/N_f$ where 
$N_b^{(0)}$ is the number of initial bosons prior to evaporation).
While our analysis confirms that the initial decrease in $T/T_F$ 
is faster for lower $\omega_f/\omega_b$ ratios, 
thus suggesting more efficient cooling - if measured by the 
drop in $T/T_F$ per unit of boson removed in the evaporation process - 
we also notice that the evaporation process extends much further 
for larger $\omega_f/\omega_b$ and allows to reach deeper Fermi
degeneracy factors before stopping. Therefore, if the goal of 
the cooling process is the achievement of the lowest $T/T_F$ 
degeneracy parameters rather than saving bosons during the 
evaporative process \cite{note3}, the use of larger $\omega_f/\omega_b$ is 
beneficial, at variance with \cite{Carr}.

The main issue in \cite{Carr} is that no temporal dependence for the 
trapping frequencies was assumed (unlike in
Eqns. (\ref{dNb}-\ref{dEf})), which is unphysical for any realistic evaporative 
cooling strategy involving species-selective trapping strengths. 
The discussion in \cite{Carr} also struggles with issues arising from
more practical limitations, as commented in \cite{BrownHayes}, 
in particular the fact that all the Fermi-Bose species available in 
practice will be affected by issues of spatial overlap in usual confining potentials, due 
to the smaller mass of the Fermi species (apart from 
the never considered ${}^{40}$K-${}^{23}$Na mixture). Also, for
$\omega_b/\omega_f > 1$ the superfluid critical velocity of 
the Bose gas will become larger than the Fermi velocity, inhibiting 
scattering between fermions and bosons and then sympathetic cooling
\cite{Chikkatur,Timmermans}.  
Finally, a weaker trapping frequency for the Fermi species corresponds
to a lower potential energy depth with respect to that of the
bosons, resulting in significant fermion losses during the forced 
evaporative process of the bosons.  As we discuss in the 
following, a large $\omega_f/\omega_b$ ratio is also beneficial 
in terms of improving the spatial overlap between fermions and bosons and 
therefore the cooling efficiency.

\begin{figure*}
\includegraphics[width=1.0\columnwidth,clip]{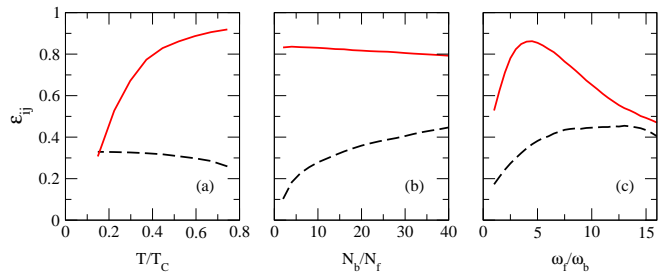}
\caption{(Color online) Overlap parameters $\epsilon_{\mbox{\tiny FB}}$ (dashed line)
and $\epsilon_{\mbox{\tiny FT}}$ (solid line) versus (a) $T/T_c$, 
(b) $N_b/N_f$, and (c) $\omega_{f}/\omega_{b}$.  Plots made with other 
two parameters held at given values of $N_{b}/N_{f}$=15,
$T/T_c$=0.3, $\omega_{f}/\omega_{b}$=3,  in the case of 
a scattering length for Rb of $a_{bb}$=+5.8 nm and an 
interspecies scattering length $a_{fb}$=+0.5 nm 
(with the same values also used for the plots in Figs. 5-7).}
\end{figure*}

\begin{figure}[t]
\includegraphics[width=1.0\columnwidth,clip]{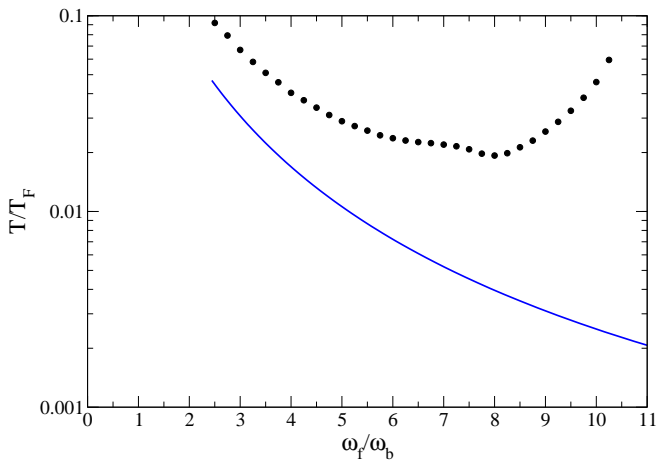}
\caption{(Color online) Plots of the minimum $T/T_F$ reached during the evaporative cooling
process in an optical dipole trap, without (line) and with (dots) the
fermion-thermal spatial overlap taken into account. The curve and dots
start at the lowest possible value of $\omega_f/\omega_b$= 2.443,
which is the `natural' trapping frequency ratio for a $^6$Li-$^{87}$Rb mixture 
with the deconfining laser beam switched off ($P_2$=0).}
\label{1DtempccTF}
\end{figure}

\begin{figure}[t]
\includegraphics[width=1.00\columnwidth,clip]{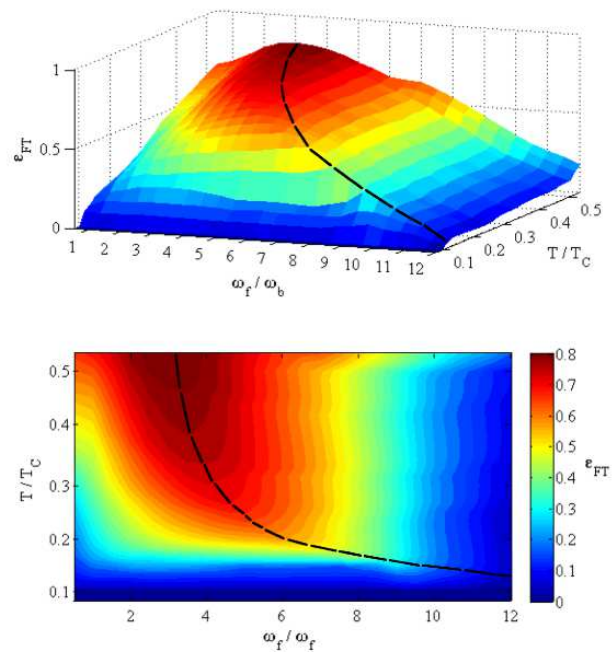}
\caption{(Color online) Three-dimensional plot and two-dimensional contour plot of
the $\epsilon_{\mbox{\tiny FT}}$ dependence on $\omega_{f}/\omega_{b}$,
and $T/T_c$.  Significant overlap values for trapping 
ratios in the range of $\omega_f/\omega_b \simeq $ 3-7 are evident at 
relatively large temperature ratios $T/T_c$. At lower temperatures, the optimal
overlap is achieved at higher trapping frequencies ratios. 
The optimal path maximizing the overlap is highlighted by 
the dashed line in both plots.}
\end{figure}

\begin{figure}[t]
\begin{center}
\includegraphics[width=0.95\columnwidth,clip]{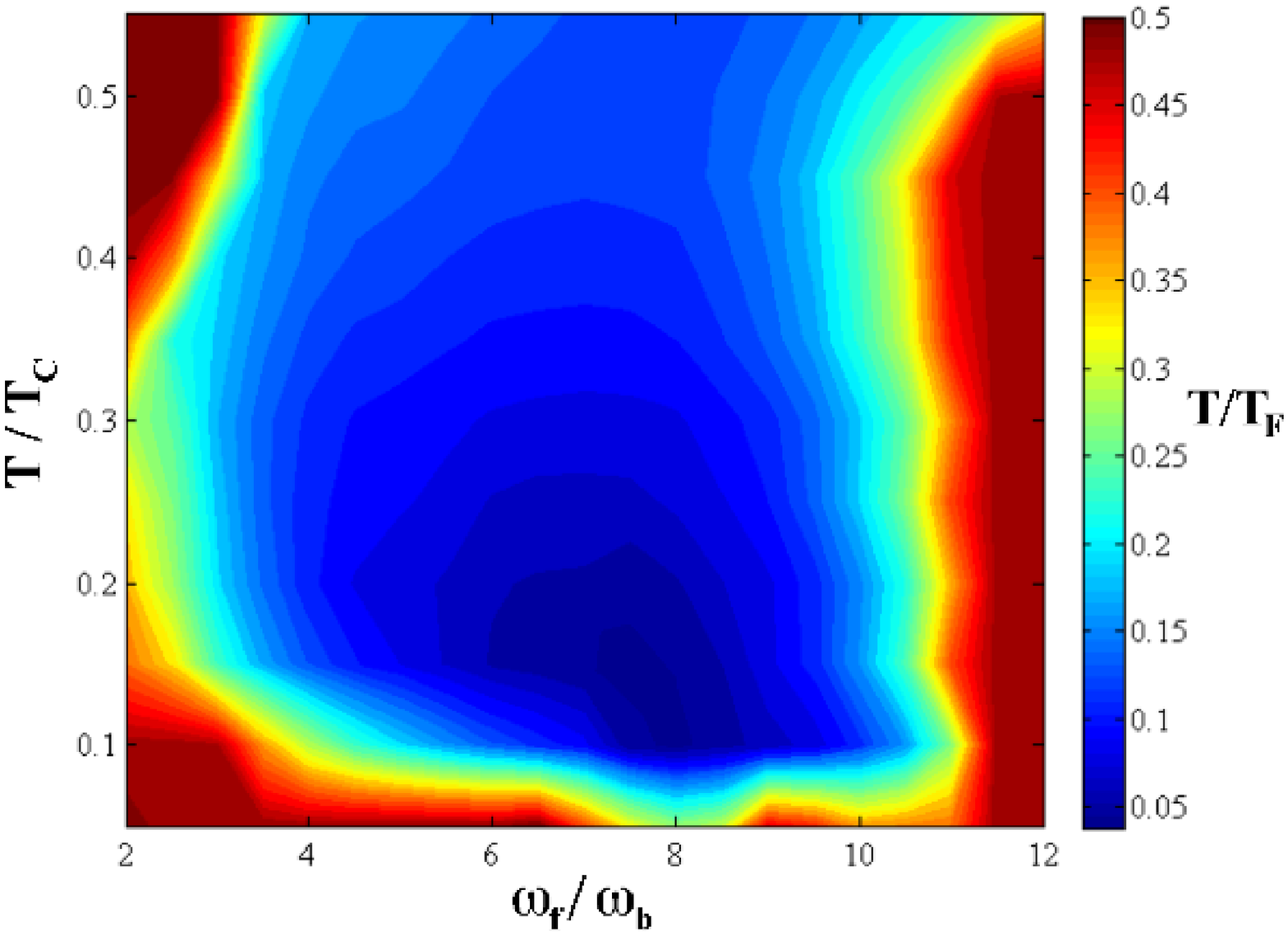}
\end{center}
\caption{(Color online) Contour plot of the Fermi degeneracy factor $T/T_F$ versus 
the Bose degeneracy factor $T/T_c$ and the trapping frequency ration 
$\omega_f/\omega_b$. The lower Fermi degeneracy factor of $T/T_F
\simeq $ 0.02 is obtained for $\omega_f/\omega_b \simeq$ 8.}
\end{figure}

\subsection{Spatial overlap between the fermion and the thermal Bose atoms}

An even more realistic analysis of the cooling dynamics must also
account for the intrinsic inhomogeneous character of the trapping 
potential as this will result in incomplete overlap between the Fermi 
and the Bose gases and consequently a decreased cooling efficiency. 
As introduced in \cite{BrownHayes}, we express the spatial overlap 
$\epsilon_{ij}$ between two clouds of densities $\rho_i$ and $\rho_j$, as
\begin{equation}
\epsilon_{ij}=(N_i N_j)^{-1/2} \int \rho_i^{1/2}(r) \rho_j^{1/2}(r) d^3 r \ ,
\end{equation}
where $i,j$ refer to the fermion (F), Bose condensate (B) or thermal
boson (T) density profiles. The fraction of atoms that share the same region
of space is thus given by $\epsilon_{ij}^2$. The Bose atoms available for 
cooling are those having a non-zero specific heat, {\it i.e.} those in a
thermal state.  Finite overlap between the thermal bosons and the 
Fermi atoms will result in a decrease in the cooling rate with 
respect to the case of ideal overlap. An accurate evaluation of the cooling
rate should take into account kinetic equations for the two
interacting gases.  In a pessimistic, conservative fashion, we can assume that the 
cooling rate $\dot{q}$ is decreased by a factor equal to the fraction of atoms 
that can actually exchange energy without any mass transport
involved, as $\dot{q}_\mathrm{cool} \rightarrow \epsilon_{\mbox{\tiny FT}}^2 \dot{q}_\mathrm{cool}$.
The minimum attainable degeneracy parameter $T/T_F$ correspondingly
increases as $T/T_F \rightarrow \epsilon_{\mbox{\tiny FT}}^{-2}~T/T_F$.
This static estimate does not take into account the timescale over
which fermions and bosons exchange energy through elastic collisions, 
and the particle relocation along the trap volume, but it can be 
considered as an upper limit to the effect of partial overlap.  This analysis requires 
the density profiles of the condensate and the non-condensed thermal boson to 
computed independently, following the discussion of a Fermi-Bose mixture 
at finite temperature reported in \cite{Dalfovo}. 

The dependence of the spatial overlap parameters
$\epsilon_{\mbox{\tiny FB}}$ and $\epsilon_{\mbox{\tiny FT}}$ on
temperature, boson number, and trapping frequency ratio is shown in
Fig. 4.  As the temperature drops below $T_c$  a finite condensate 
fraction appears;  the fermion-thermal boson overlap starts to decrease, 
and the fermion-condensate overlap increases.  
There is a limit to $\epsilon_{\mbox{\tiny FB}}$ for a
given $\omega_f/\omega_b$, however, since the cloud of less massive fermions 
will have a much larger spatial radius and thus a strong relative confinement 
is required to improve the fermion-Bose condensate overlap.  
The overlap dependence on number ratio is
rather straightforward, with a gradual decrease in fermion-condensate
overlap as bosons are evaporated and an almost flat behaviour for
$\epsilon_{\mbox{\tiny FT}}$ since $N_b^{ex}$ is roughly constant for a
given temperature and boson losses manifest as a decrease in the
condensate fraction and have minimal effect on the thermal cloud.  

The effect of the spatial overlap on the minimum reachable Fermi 
degeneracy factor $T/T_F$ is depicted in Fig. 5, with the choice of 
interspecies scattering length of the Li-Rb system corresponding to
the pessimistic scenario of repulsive interaction (see \cite{Marzok} 
for the issues related to its measurement). 
It is evident that about one order of magnitude may be lost in the
achievable minimum degeneracy factor when the overlap factor is taken into account,
although we conjecture that with a full kinetic analysis the 
actual result will be located in between the two curves. 
The situation is analyzed in more detail in Fig. 6, which shows 
$\epsilon_{\mbox{\tiny FT}}$ versus $\omega_{f}/\omega_{b}$ and
$T/T_c$. During the cooling process, {\it i.e.} as 
$T/T_c$ decreases, the optimal overlap is shifted to larger values of 
$\omega_f/\omega_b$, until the minimum of the Fermi degeneracy is 
reached as shown in the contour plot of Fig. 7. 
This suggests the use of a time-variable trapping strategy, initial with lower 
values of $\omega_f/\omega_b$, then increased in time by increasing  
the power ratio $P_2/P_1$.
Such a time-dependent relative confinement strategy is not the only way to 
optimize the Fermi degeneracy factor, however. In the following, we 
will discuss a similar procedure which exploits the advantages 
of reducing the dimensionality of the system when Fermi degeneracy 
is approached.

\section{Heat capacity matching through lower dimensionality}

As an alternative to the cooling strategy described above, we discuss
here the possibility to match the heat capacity of Bose and Fermi
gases at the lowest possible $T/T_F$ by exploiting lower 
dimensionality traps. Ultimately, the mismatching between the 
specific heats of Bose and Fermi degenerate gases depend on 
the scaling of the heat capacities with temperature, and 
this in turn depends upon the dimensionality of the Bose gas.
As discussed in \cite{RehrPRB1,BagnatoPRA44} and demonstrated in 
\cite{KetterlePRA54,MoritzPRL94}, a dramatic increase in the 
trapping frequency in one (or two) trapping axes will result in
an effective two- (or one-) dimensional system.  
This in turn allows for a better matching of the heat capacities 
since the Bose gas dependence on temperature will become milder 
than in the full 3D case. In order to gain quantitative insights on
how to realize such a matching, we first consider noninteracting gases 
in a harmonic potential, with the number $N_{f,b}$ of particles fixed.  

The total number of fermions and bosons is evaluated as ($+$ for 
fermions, $-$ for bosons):
\begin{equation}
N_{f,b} (\mu, T) = \sum_{j=0}^{\infty} \frac{g_j}
{e^{(E_j - \mu)/k_B T} \pm 1}_{  \ \  ,}
\label{eq:numbers}
\end{equation}
\noindent
where $g_{j}$ is the degeneracy of energy level $E_{j}$, $\mu$ the
chemical potential, $k_{B}$ Boltzmann constant, and $T$ the
temperature, with the number of particles $N_{f,b}$ fixed. 
Solving this equation numerically for $\mu = \mu (T)$, we can 
then calculate the total energy
\begin{equation}
E_{f,b} (T) = \sum_{j=0}^{\infty} \frac{g_j E_j}{e^{(E_j - \mu(T))/k_B T} \pm 1}_{ \ \ ,}
\label{eq:energies}
\end{equation}
\noindent
and from this we obtain the heat capacity as $C (T) = \partial E/\partial T$.
In the calculation below, we assign $N_{f,b}=10^4$, and assume the
initial trap to be isotropic with the trapping frequency $\omega =
2\pi \times 15.87$ kHz.  The numerical calculations for 2D and 3D 
are straightforward, while for 1D some approximations are necessary 
to reduce the simulation time to realistic values.

\begin{figure}[t]
\begin{center}
\includegraphics[width=1.00\columnwidth,clip]{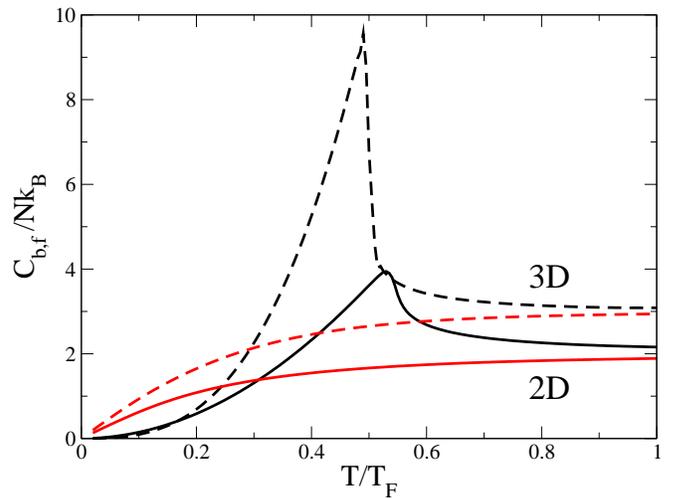}
\end{center}
\caption{(Color online) Heat capacity curves of bosons and fermions in 3D (dashed
curves) and 2D (solid curves) in an isotropic harmonic trap, with the
bosons exhibiting nonmonotonic behaviour. The crossing point between 
the Bose and Fermi curves for the 2D case occurs at a slightly higher 
$T/T_F$ value, ruling out its use for a more favourable cooling of fermions.}
\end{figure}

\subsection{2D and 3D traps}

For atoms trapped in a three dimensional harmonic potential 
$V (r) =m\omega^2r^2/2$, the energy eigenvalues $E_j$
($j$=0,1,2,\ldots) are given by $E_j^{3D} = (j + 3/2) \hbar \omega$.
Since the trap is three-dimensional and isotropic, the degeneracy
$g_j$ of the energy levels is given by $g_j = (j+1)(j+2)/2$.  
The number of particles for bosons ($-$) and fermions ($+$) is:
\begin{equation}
N_{f,b}^{3D} = \frac{1}{2}\sum_{j=0}^{Q} \frac{(j +1 )(j + 2)} 
{e^{\left((j + 3/2) \hbar \omega - \mu\right)/k_B T} \pm 1}_{ \ \ .}
\label{eq:N3D}
\end{equation}
The upper limit $Q$ in the summation should be infinity in principle,
but a value of $Q$=1500 is sufficient for numerical convergence. 
For the given parameters $N_{f,b}$ and $\omega$ we solve the above two
equations for $\mu$ at different temperatures, and then calculate the
heat capacities.  As depicted in Fig. 8, the Bose and Fermi heat
capacities intersect each other at $T \simeq 0.293 T_F$, with
the three-dimensional Fermi temperature $T_F^{3D} = (6N_f)^{1/3} \hbar
\omega /k_B$. The situation is very similar for atoms trapped in a 2D isotropic
harmonic potential, except that now the energy eigenvalues are given
by$E_j^{2D} = (j+1)\hbar \omega$, with degeneracy $g_j=j+1$, and
$Q$ needs to be increased to $\sim 10^5$ to achieve adequate
convergence. The two heat capacities curves intersect each other at $T
\simeq 0.308~T_F$ where the 2D Fermi temperature $T_F$ is given by
$T_F^{2D} = (2N_f)^{1/2} \hbar \omega/k_B$.
From Fig. 8 we see that going from a full 3D to a 2D system actually 
slightly worsens the heat capacity matching, yielding a higher $T/T_F$ at 
the point where $C_{b}$ and $C_{f}$ intersect each other. 
However, further reduction to a 1D system results in complete matching of 
the heat capacities, as we will see below.  

\begin{figure}[t]
\begin{center}
\includegraphics[width=1.00\columnwidth,clip]{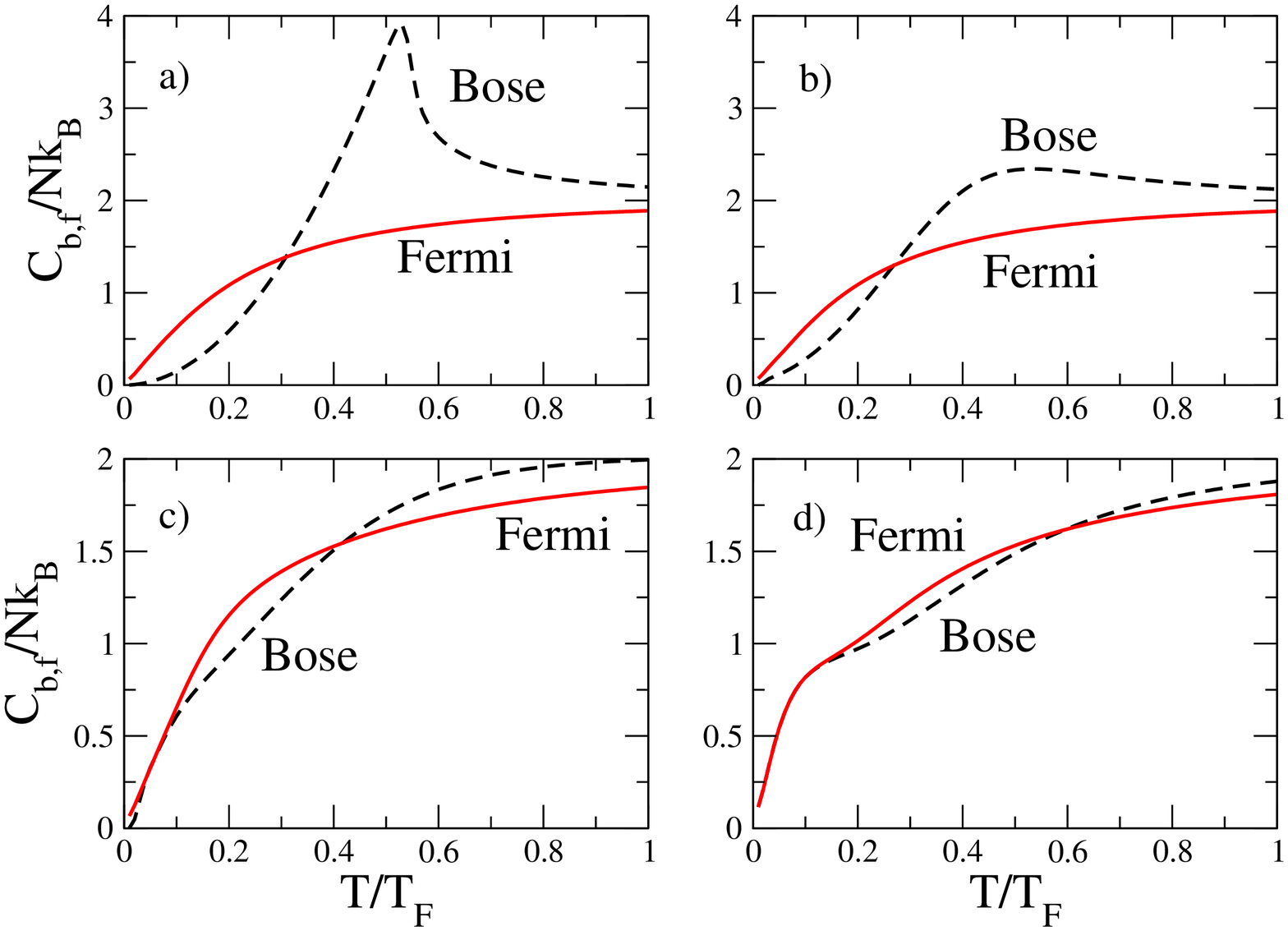}
\end{center}
\caption{(Color online) Heat capacity curves of bosons 
(black, dashed) and fermions (red, continuous). 
We consider for simplicity the case of equal number of bosons and
fermions $N_b = N_f= 10^4$ with equal mass $m_b = m_f$. 
Curves in plots (a), (b), (c) and (d) are in 2D traps with $k =
\omega_y/\omega_x$ equals to $1$, $2.5 \times 10^3$,  $2 \times
10^4$, and $5 \times 10^4$ respectively, with the case in (d) 
showing an initial plateau of the heat capacities at 
$\simeq 1$ before reaching the high temperature limit of a 
2 D gas.}
\end{figure}

\subsection {1D trap}

For atoms in a 1D trap, the energy eigenvalues are given by $E_j^{1D}
=(j+1/2)\hbar \omega$, with degeneracy $g_j$ = 1. If we follow the
same steps as the 2D and 3D cases, the upper summation limit $Q$ (see
Eq. (\ref{eq:N3D})) must be extremely large in
order to reach convergence and approximations are required for the
one-dimensional trap analysis.  These approximations are outlined in 
Appendix B and with them we obtain the Fermi and Bose energies, from 
which the heat capacities are evaluated. To take full advantage of 
the gain in the heat capacity matching in 1D
traps it is helpful to investigate how heat capacities of bosons and
fermions change when we gradually reduce the dimension of the trap. 
We assume that the atoms are first trapped in a 2D trap with trapping
frequencies along the two dimensions $\omega$ and $k \omega$, and then 
this relative confinement parameter 
$k$ is gradually increased from 1 to infinity. The system will be
effectively 1D when $k_B T \ll k \hbar \omega$. We evaluate the heat 
capacities of bosons and fermions at different values of $k$. 
The results are shown in Fig. 9 (a-d). As $k$ increases, the
shape of the heat capacity curve of bosons becomes more similar to
that of fermions, in the sense that it slowly loses the peak structure  
and the curvature near zero temperature begins to resemble that 
of fermions. If $k$ is further increased (case (c))
there will be a region where the two curves completely coincide with each other. 
This is consistent with the previous result we obtained for an ideal 1D
trap. In Fig. 9 (d)  the heat capacity curves of bosons 
and fermions are shown at an even higher aspect ratio, to emphasize
the crossover from the 1 D case to the 2 D case at high temperatures.
In the ideal 1D case, the heat capacity curves are identical, as simply explained
in the canonical ensemble approach \cite{1Densemble1,1Densemble2}. 
Indeed, the total internal energy of fermions in a 1D harmonic trap 
only differs from that of the bosons by the Fermi zero-point energy 
$E_0 =N_f(N_f-1)\hbar \omega/2$, and therefore the two systems have 
identical heat capacities.

The existence of this crossover indicates that we can control 
the heat capacity matching of bosons and fermions by changing 
the ratio of the two trapping frequencies in a 2D trap. 
Thus one possible solution to improve the cooling efficiency is to
first evaporate in a 3D trap and then, when the Fermi degeneracy
starts to reach about $T/T_F \simeq 0.3$, to increase the trapping 
frequencies achieving a quasi-one dimensional system, then continuing 
the evaporation process.
A possible limitation of this technique comes from the larger 
collisional loss rate as a result of the increased confinement.
Also, as studied for achieving Bose condensation of hydrogen atoms, 
the nearly 1D character of the evaporative cooling \cite{Pinkse} may 
lead to non-ergodic evaporation limiting its efficiency \cite{Surkov},
although this has not prevented achievement of Bose degeneracy \cite{OneDexp}.

\section{Conclusions}

We have examined the thermodynamics of evaporative and sympathetic
cooling in a Fermi-Bose mixture, and identified possible ways to achieve a lower
Fermi degeneracy factor $T/T_F$. 
Thermodynamical considerations are based on general assumptions and
measurable, phenomenological inputs, like heating rate and specific 
heat, and provide a solid framework to discuss cooling dynamics regardless 
of sophisticated microscopic models \cite{Blakie}. 
They also allow for comparison with experimental results, such as
those discussed in \cite{Raman,Kinast,Gerbier,Gati1,Gati2,Luo}, or 
for the inclusion of more realistic inputs such as the specific 
heat of an interacting gas \cite{Giorgini}.
Although we have focused the attention on the particular
$^6$Li-${}^{87}$Rb mixture, the extension to other Fermi-Bose
combinations is straightforward, furthermore benefiting from more  
favourable interspecies thermalization properties with respect to 
this particular mixture, for which limitations in elastic scattering 
and use of Feshbach resonances have been experimentally evidenced \cite{Silber,Deh}.

Among the main results we have obtained, we have discussed two different cooling 
strategies: constructing a species-selective trap with independently tunable 
Fermi and Bose trapping frequencies, and creating traps with reduced 
dimensionality in the latest stage of evaporation. We have shown 
that different trapping ratios lead to distinctly different cooling trajectories.   
However, incomplete spatial overlap will not only result
in a longer cooling time needed to attain a given temperature, but
will also increase the temperature at with the heating rate will
balance the cooling rate. When the progressive depletion of the
bosonic thermal cloud is taken into account, optimized cooling
requires time-dependent trapping strengths. Additionally, we 
have discussed how to exploit the strong dependence of the
bosonic specific heat upon dimensionality to create nearly 
one-dimensional traps in the ultimate stage of sympathetic cooling.  
This will be of particular relevance for various planned studies 
of Fermi gases in optical lattices \cite{Opticalattice} in which
bichromatic optical traps are not viable. 

\appendix
\section{Bose and Fermi gases in a 3D harmonic trap of finite depth}
\label{appendix1}

We report here the expressions for the energy and the number of particles
of ideal degenerate Bose and Fermi gases confined a in a three-dimensional 
harmonic trap of finite depth.

For the trapped bosons we start with the exact expressions of the
energy and of the number of particles in a general 3D harmonic potential:
\begin{equation}
E_b = \sum_{n_x=0}^\infty\sum_{n_y=0}^\infty\sum_{n_z=0}^\infty
\frac{E_{n_x,n_y,n_z}}{e^{(E_{n_x,n_y,n_z}-\mu)/k_BT}-1},
\label{Eb1}
\end{equation}
\begin{equation}
N_b = \sum_{n_x=0}^\infty\sum_{n_y=0}^\infty\sum_{n_z=0}^\infty \frac{1}
{e^{(E_{n_x,n_y,n_z}-\mu)/k_BT}-1},
\label{Nb1}
\end{equation}
where $E_{n_x,n_y,n_z}=\hbar\omega_{bx}(n_x+1/2)+\hbar\omega_{by}(n_y+1/2)
+\hbar\omega_{bz}(n_z+1/2)$. As usual, we write $N_b=N_b^0+N_b^{ex}$,
where $N_b^{0}=(e^{E_{0,0,0}/k_BT}-1)^{-1}$ is the number of bosons in 
the ground state and $N_b^{ex}$ the number of those thermally excited.
In order that the number of particles remains positive, it is necessary for
the chemical potential to satisfy $\mu(T) \leq E_{0,0,0}$.
At temperatures $T\leq T_c$, where $T_c$ is the Bose-Einstein condensation
critical temperature, the chemical potential is frozen to its maximum value.
In general, $\mu(T)$ and all the other thermodynamic quantities can be 
evaluated explicitly as a power series expansion in the two parameters
$\mu(T) - E_{0,0,0}$ and $\hbar\omega_b/k_BT$, where 
$\omega_b=(\omega_{bx}\omega_{by}\omega_{bz})^{1/3}$ \cite{Kirsten}.
In the case $T \lesssim T_c$ and $\hbar\omega_b \ll k_BT$, 
which is relevant to the experimental situations discussed here, 
we can restrict to the lowest order and write
\begin{eqnarray}
E_b &=& 3 \zeta(4) \frac{(k_BT)^4}{(\hbar\omega_b)^3},
\label{Eb2}
\end{eqnarray}
\begin{eqnarray}
N_b^{ex} &=& \zeta(3) \left(\frac{k_BT}{\hbar\omega_b}\right)^3,
\label{Nb2}
\end{eqnarray}
where $\zeta$ is the Riemann zeta function.
Alternatively, the above two results can be obtained using the semiclassical density of states
\begin{eqnarray}
\varrho_b(E) = \frac{d}{d E} \frac{\frac16 E^3}
{\hbar\omega_{bx} \hbar\omega_{by} \hbar\omega_{bz}} =\frac{E^2}{2 (\hbar\omega_{b})^3}
\label{g}
\end{eqnarray}
and the continuum approximation 
\begin{eqnarray}
E_b &=& \int_0^{\infty} \frac{E}{e^{E/k_BT}-1} \varrho_b(E) dE,
\label{Eb3}
\end{eqnarray}
\begin{eqnarray}
N_b^{ex} &=& \int_0^{\infty} \frac{1}{e^{E/k_BT}-1} \varrho_b(E) dE.
\label{Nb3}
\end{eqnarray}
For a trap of finite depth schematized as a harmonic potential truncated
at energy $\Delta U_b$, we thus write
\begin{eqnarray}
E_b &=& \int_0^{\Delta U_b} \frac{E}{e^{E/k_BT}-1} \varrho_b(E) dE
\nonumber \\ 
&=& \frac{(k_BT)^4}{(\hbar\omega_b)^3} 
\frac{1}{2} \int_0^{\frac{\Delta U_b}{k_BT}} \frac{t^3}{e^{t}-1} dt,
\label{Eb4}
\end{eqnarray}
\begin{eqnarray}
N_b^{ex} &=& \int_0^{\Delta U_b} \frac{1}{e^{E/k_BT}-1} \varrho_b(E) dE
\nonumber \\
&=& \frac{(k_BT)^3}{(\hbar\omega_b)^3} 
\frac{1}{2} \int_0^{\frac{\Delta U_b}{k_BT}} \frac{t^2}{e^{t}-1} dt.
\label{Nb4}
\end{eqnarray}

Consider now a system of $N_f$ fermions confined by a harmonic trap
having characteristic frequencies $\omega_{fx},\omega_{fy},\omega_{fz}$.  
Under the condition
$\hbar\omega_{fx},\hbar\omega_{fy},\hbar\omega_{fz}\ll k_{B}T$,
a continuum approximation holds as in the boson case, so that
the fermion counterparts of Eqs. (\ref{Eb1}) and (\ref{Nb1}) can be 
simplified to
\begin{eqnarray}
E_f &=& \int_0^\infty \frac{E}{e^{(E-\mu)/k_BT}+1}\varrho_f(E)dE,
\label{Ef1}
\end{eqnarray}
\begin{eqnarray}  
N_f &=& \int_0^\infty \frac{1}{e^{(E-\mu)/k_BT}+1} \varrho_f(E)dE,
\label{Nf1}
\end{eqnarray}
where $\varrho_f(E)=E^2/2(\hbar\omega_f)^3$
and $\omega_f=(\omega_{fx}\omega_{fy}\omega_{fz})^{1/3}$.
Note that in the above expressions we have neglected the zero-point energy
which is justified since for fermions $\mu \gg E_{0,0,0}$.
The accuracy of the continuum approximation introduced above is
discussed in detail in \cite{Toms}.
The chemical potential $\mu(T)$ can be eliminated between Eqs. (\ref{Ef1}) 
and (\ref{Nf1}) by means of the standard Sommerfeld expansion in powers of
temperature. 
By keeping terms only to second order in $T$, we have the well known result
\begin{eqnarray}
E_f(N_f,T) \simeq E_f(N_f,0) + \frac{\pi^2}{6}(k_BT)^2 \varrho(E_F),
\label{Ef2}
\end{eqnarray}
where the Fermi energy $E_F$ is related to $N_f$ by 
\begin{eqnarray}
N_f = \int_0^{E_F} \varrho_f(E) dE= \frac{1}{6}
\left(\frac{E_F}{\hbar \omega_f}\right)^3
\label{Nf2}
\end{eqnarray}
and the zero temperature term $E_f(N_f,0)$ is given by 
\begin{eqnarray}
E_f(N_f,0) = \int_0^{E_F} E \varrho_f(E) dE= 
\frac{{E_F}^4}{8 (\hbar \omega_f)^3}.
\label{Ef3}
\end{eqnarray}
Making use of the relationship between the Fermi energy and 
the number of fermions, we finally obtain
\begin{equation}
E_f=\frac{3}{4}6^{1/3} N_f^{4/3} \hbar \omega_f +
\frac{\pi^2}{2}  6^{-1/3}N_f^{2/3} \frac{(k_B T)^2}{\hbar \omega_f}.
\label{Ef4}
\end{equation}
Equations (\ref{Nf2}) and (\ref{Ef4}) are valid also for fermions trapped
into a harmonic potential truncated at energy $\Delta U_f$, provided that
$\Delta U_f \gg E_F, k_B T$.

\section{Internal energy for a 1D trapped ideal Fermi gas}

The total atom numbers in one dimension $N_b^{1D}$ and $N_f^{1D}$ are
given by Eq. (\ref{eq:numbers}), with $g_j=1$.  However, the upper
summation limit $Q$ (see Eqs.  (\ref{eq:N3D}))
needed for convergence becomes unreasonably large in one dimension 
\cite{dimensions}, and thus approximations are required.

For fermions, we can directly replace the summation with an integral
in the case of $k_B T \gg \hbar \omega$:
\begin{eqnarray}
N_{f}^{1D} &=& \sum_{j=0}^{\infty}\frac{1}{e^{\left((j + 1/2)\hbar \omega- \mu\right)/k_B T} + 1} 
\nonumber \\ 
& \simeq & \frac{1}{\hbar \omega} \int_{\hbar \omega /2}^{+\infty}\frac{dE}
{e^{\left(E - \mu\right)/k_B T} + 1}.
\label{eq:Nf1D}
\end{eqnarray}

Similarly, the total energy is given by
\begin{equation}
 E_f \simeq \frac{1}{\hbar \omega} \int_{\hbar\omega/2}^{+\infty}\frac{EdE}{e^{(E - \mu)/k_B T} + 1}.
\label{eq:Uf1D}
\end{equation}

The total number of bosons is given by
\begin{eqnarray}
N_{b}^{1D} & = & \sum_{j=0}^{+\infty} \frac{1}{e^{\left((j + 1/2) \hbar \omega - \mu\right)/k_B T} - 1} \nonumber\\
&   &  = \frac{1}{e^{\left(\hbar \omega /2 - \mu\right)/k_B T} - 1} \nonumber \\
&   &  + \sum_{j=1}^{+\infty} \frac{1}{e^{\left((j+1/2) \hbar \omega - \mu\right)/k_B T} - 1} {\ .}
\label{eq:Nb1D}
\end{eqnarray}
The second summation term can be evaluated by integral using the
Euler-Maclaurin formula:

\begin{eqnarray*}
\sum_{x=a}^b F(x) = 
& &    \int_a^bF(x)dx + \frac{F(a)}{2} + \frac{F(b)}{2}
\noindent \\
& & + \sum_{k=1}^n \frac{B_{2k}}{(2k)!} [F^{(2k-1)}(b) - F^{(2k-1)}(a)] + R
\label{eq:EMequ}
\end{eqnarray*}

\noindent
where $B_2$ = 1/6, $B_4$ = -1/30,\ldots are the Bernoulli numbers, and
$R$ is the remainder term.  In the case of
$k_{B}T \gg \hbar \omega$, the first term ($k$ = 1) in the expansion
is sufficient, giving:

\begin{eqnarray}
N_{b}^{1D}& = & \frac{1}{e^{\left(\hbar\omega/2-\mu\right)/k_B T}-1} \nonumber \\
& + & \int_1^{+\infty}\frac{dx}{e^{\left[(x + 1/2)\hbar\omega-\mu\right]/k_B T}-1} \nonumber \\
& + & \frac{1}{2\left[e^{\left(3\hbar\omega/2-\mu\right)/k_BT}-1\right]} \nonumber \\
& - & \frac{\hbar \omega}{12 K_B T} 
\frac{e^{\left(3\hbar\omega/2-\mu\right)/k_B T}}{{\left(e^{(3\hbar\omega/2-\mu)/k_B T}-1\right)}^2} {\ .}
\label{eq:Nb1DEM}
\end{eqnarray}

Having fixed N, we solve this equation numerically for $\mu$ at
different temperatures to obtain $\mu = \mu (T)$, and then apply the
Euler-Maclaurin formula again to obtain the total energy $E_b^{1D}$.

The discussion above holds for a one dimensional Bose gas. In the 
case of a trapping frequency range for which there is a smooth
crossover from two dimensions to one dimension, we recall that 
the number of particles in a 2D trap is given by: 

\begin{equation}\label{eq:N2d0}
N_{f,b}^{2D} = \sum_{i,j=0}^{+\infty} \frac{1}{e^{\left[(i + 1/2) 
\hbar \omega_1 + (j + 1/2) \hbar \omega_2 - \mu\right]/k_B T} \pm 1}.
\end{equation}

Running two independent indexes is computationally very inefficient
and slow. We therefore assume $\omega_1 = \omega$ and $\omega_2 = k
\omega$ where $k$ is a positive integer. The summation is run until 
$(i + k j) = Q$ which, with $Q$ large enough, yields a good approximation. 
Now $N_{f,b}^{2D}$ becomes:
\begin{equation}\label{eq:N2d1}
\begin{split}
N_{2d}& = \sum_{i,j=0}^{i+kj=Q} \frac{1}{e^{\left[(i + k j) 
\hbar \omega + (1/2 + k/2) \hbar \omega - \mu\right]/k_B T} \pm 1}\\
      & = \sum_{i=0}^{Q} \frac{floor(i/k) + 1}
{e^{\left[i \hbar \omega + (1/2 + k/2) \hbar \omega - \mu\right]/k_B T} \pm 1},
\end{split}
\end{equation}
where $floor(j/k)$ is the nearest integer less than or equal to
$j/k$. In this way only one index is present, allowing to improve
significantly the computational speed.

\begin{acknowledgments}
MBH and QW acknowledge support from the Dartmouth Graduate Fellowship
program, and MBH also acknowledges support from the Gordon Hull and 
NSF-GAANN fellowships.  
CP and RO acknowledge partial support through Cofinanziamento MIUR 
protocollo 2002027798$\_$001, and RO also acknowledge partial support 
by the NSF through the Institute for Theoretical Atomic and Molecular 
Physics at Harvard University and the Smithsonian Astrophysical Observatory.
\end{acknowledgments}


\begin{thebibliography}{99}
\bibitem{note2} There is a trade-off in decreasing the boson trapping 
frequency $\omega_b$ since the efficiency of evaporative cooling in 
itself is inversely proportional to the boson heat capacity. 
Also, a smaller $\omega_b$ implies lower densities of the Bose cloud, 
implying slower thermalization. Cooling strategies in which the boson 
trapping frequency is instead kept constant, while the fermion
trapping frequency is increased, are possible alternatives to 
overcome these limitations.
\bibitem{note3} 
It is worth to point out that in most of the experiments using 
Fermi-Bose mixtures and focusing on fermion studies, the residual 
Bose component is anyway removed after sympathetic cooling.
\end{thebibliography}
\end{document}